\documentclass[twocolumn,showpacs,preprintnumbers,amsmath,amssymb]{revtex4}

\usepackage{graphicx}
\usepackage{dcolumn}
\usepackage{bm}

\begin{document}

\preprint{accepted for publication in Phys. Rev. Lett.}

\title{STS Observations of Landau Levels at Graphite Surfaces}

\author{T. Matsui, H. Kambara, Y. Niimi, K. Tagami, M. Tsukada, and Hiroshi Fukuyama$^{\ast}$}

\affiliation{Department of Physics, Graduate School of Science, University of Tokyo, 7-3-1 Hongo, Bunkyo-ku, Tokyo 113-0033, Japan}

\date{April 13, 2005}

\begin{abstract}
Scanning tunneling spectroscopy (STS) measurements were made
on surfaces of two
different kinds of graphite samples, Kish graphite and highly oriented
pyrolytic graphite (HOPG), at very low temperatures and in high magnetic
fields.
We observed a series of peaks in the tunnel spectra associated with Landau
quantization of the quasi two-dimensional electrons and holes.
Comparison with calculated local density of states at the surface layers
allows us to identify Kish graphite as bulk graphite and HOPG as graphite
with finite thickness of 40 layers.
This explains the qualitative difference between the two graphites
reported in the recent transport measurements which suggested the quantum
Hall effect in HOPG.
This work demonstrates how powerful the combined approach between the high 
quality STS measurement and the first-principles calculation is in material 
science.
\end{abstract}

\maketitle

Graphite is one of the best studied materials concerning its
electronic properties both experimentally and theoretically.
It is a semimetal or zero gap semiconductor with a unique \textit{massless}
linear dispersion relation for the conduction band.
Graphite is also an important mother system for technologically attractive
materials such as carbon nanotubes, fullerene, nanographites, etc.
Thus, any new insight on this material should have broad impacts.

Recently, graphite has been attracting renewed fundamental interests
particularly for its transport properties in magnetic fields.
Kopelevich and coworkers \cite{kopelevich} observed a plateau in the Hall
resistance for highly oriented pyrolytic graphite (HOPG) in the quasi
quantum limit, which suggests the quantum-Hall effect (QHE) characteristic
of the 2D electron systems (2DES) with high mobility.
But this behaviour was not observed for a Kish graphite sample
\cite{kopelevich}.
Although the QHE is theoretically predicted for a 2D graphene sheet
\cite{zheng}, graphite is a far less ideal 2D conductor in spite of its
layered structure.
Reentrant field-driven metal-insulator transitions were also reported
\cite{kopelevich,kempa} for both types of graphites.
Theories \cite{khveshchenko} predict the electron-hole pairing
as a possible mechanism.
Even for zero field properties, the gapless spin-1 neutral collective mode
has been recently suggested from the view point of the resonating
valence bond \cite{baskaran}.

The aim of this work is to investigate the electronic states of Kish
graphite and HOPG in high fields and low temperatures by scanning tunneling
spectroscopy (STS).
The STS technique, which probes the local density of states (LDOS), should
potentially be a powerful tool to study the Landau quantization in 2DES in
magnetic fields.
This was recently demonstrated for the adsorbate-induced 2DES at the
surface of InAs(110) with evaporated Fe submonolayers \cite{morgenstern}.
For both types of graphites, we observed clear peak structures in the
tunnel spectra associated with the Landau levels (LLs).
This is the first STS observation of the LLs in bulk materials.
Comparisons with our LDOS calculations reveal that HOPG has a rather
distinct 2D electronic state due to the effective finite thickness but Kish
graphite does not.
This provides a spectroscopic explanation for the qualitatively
different transport properties between the two types of graphites and the
possible QHE in HOPG.
More generally, it is an important achievement that in combination with
the theoretical calculations we could determine the surface and
underneath electronic states in layered materials in
quantitative manners from the STS measurements.

The STS data were taken with a recently constructed ultra-low temperature
STM \cite{ULT-STM} at temperatures below 50 mK and in magnetic fields up to
6 T.
The Kish graphite sample \cite{kish} is single crystalline graphite grown
as precipitation from molten iron.
The HOPG sample \cite{hopg} is synthesized by chemical vapor deposition and
subsequent heat treatment under high pressures.
HOPG is polycrystalline graphite with ordered $c$-axis orientation.
The typical microcrystallite (platelet) size is of the order of 1 $\mu$m.
The samples were cleaved in air and then quickly loaded into an ultra high
vacuum chamber ($P < 2\times10^{-8}$ Pa) of the STM.
The differential tunnel conductance ($dI/dV$), which is proportional to the
sample density of states (DOS), was measured by the lock-in technique with
a bias modulation of $V_{{\rm mod}} =$ 0.6 or 1.4 mV at a frequency of 131
or 877 Hz.
Here $V$ is the bias voltage applied to the sample relative to the tip.

\begin{figure}
\includegraphics[width = 6.6 cm]{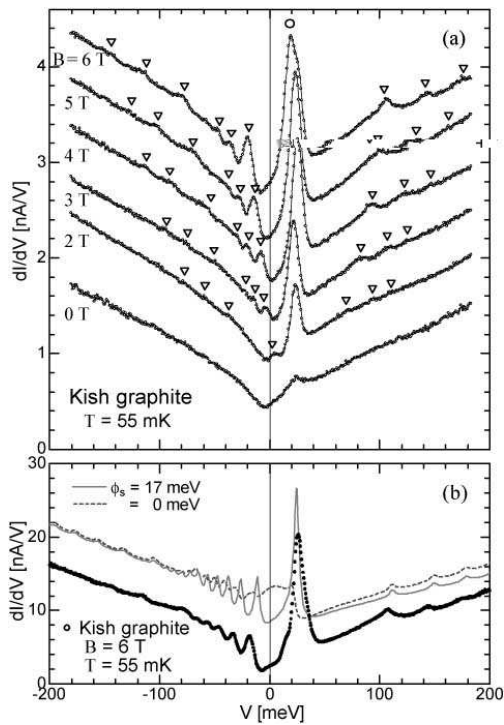}
\caption{\label{fig:kish} (a) Tunnel spectra measured at $T = 55$ mK for
Kish graphite in several different fields applied perpendicular to the
basal plane ($V= 0.2$ V, $I= 0.24$ nA). Each spectrum is vertically shifted
by 0.5 nA/V for clarity.
(b) Comparisons between the calculated LDOS ($\eta = 5$ meV) for the
surface layer of bulk graphite with (solid line) and without (dashed line)
the surface potential ($\phi_{s} = +17$ meV) and the measured tunnel
spectrum for Kish graphite at $B=6$ T (dots; $V= 20$ mV, $I$ = 0.1 nA).}
\end{figure}

Fig. \ref{fig:kish}(a) shows measured tunnel spectra for Kish graphite at 
several different fields.
Each spectrum at a given field is an average of raw data taken at 400
different positions over a $2 \times 2$ nm$^2$ area which is much smaller
than the platelet size.
The ``V" shaped spectrum with a non-zero DOS at $V =$ 0, i.e., $E = E_F$,
in zero magnetic field is characteristic of semimetallic properties of
graphite.
To our knowledge, the DOS of graphite has never been measured with such a
high energy resolution before.

In finite fields ($B$), several peaks appear superimposed onto the V-shaped
base line (the open triangles in Fig. \ref{fig:kish}(a)).
The peaks become more pronounced at higher fields, and the peak energies
are roughly in proportion to $B$.
A typical energy separation between the two successive peaks is 20-40 meV
at $B =$ 6 T which is in the same order as $\omega_{c}$ for the electrons
($= 12$ meV) and holes ($= 18$ meV) calculated from the in-plane effective
masses ($m_{e}^{*} = 0.057m_{e}$, $m_{h}^{*} = 0.039m_{e}$). Here
$\omega_{c} = eB/m^{*}$ is the cyclotron frequency and $m_e$ is the bare 
electron mass.
Thus we believe that these peaks are associated with the LLs in graphite.
In addition to the field dependent small peaks, there exists a much
pronounced peak just above $E_F$ ($V = +20$ mV) for each spectrum.
The peak energy is nearly field independent, but the amplitude grows
rapidly with increasing field.
The absence of field dependence of the peak energy strongly suggests that
it originates from the $n = 0$ (electron) and $-1$ (hole) LLs \cite{nakao}.
These levels are characteristic of the graphite lattice structure
\cite{nakao,dresselhaus_trigonal} and have been known so far only
indirectly.
The important point is that this peak almost disappears in zero field.

\begin{figure}
\includegraphics[width = 6.6 cm]{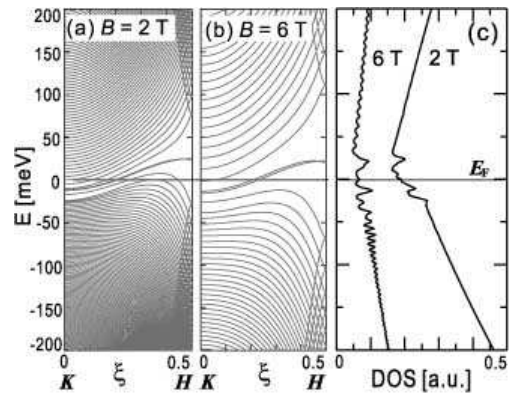}
\caption{\label{fig:bulk_band} Calculated Landau subbands along the graphite
Brillouin zone edge (K-H axis) at (a) $B=2$ T and (b) 6 T. (c) The density
of states calculated from the subbands at $B=2$ and 6 T with $\eta = 5$
meV.}
\end{figure}

Let us now compare the observed tunnel spectra with theoretically expected
LLs in bulk graphite.
Figs. \ref{fig:bulk_band}(a)(b) show Landau subbands calculated from the 
SWMcC model \cite{SWMcC} along the Brillouin zone edge (K-H axis) at $B =$ 
2 and 6 T.
We followed here Nakao's method \cite{nakao} who avoided the infinite
dimension problem of the magnetic Hamiltonian matrix by grouping it into
three submatrices and then by truncating them at finite dimensions.
We truncated them at 240 and 60 dimensions for $B = 2$ and 6 T
respectively, which are sufficiently large for studying the DOS in the
energy range interested here.
The seven band parameters in the SWMcC model were fixed to those used in
Ref.~\cite{nakao}, and small Zeeman energies are neglected throughout this
Letter.
As seen in Fig. \ref{fig:bulk_band}(b), the doubly degenerated narrow bands 
across $E_F$,
i.e., the lowest order LLs ($n = 0, -1$), are nearly unchanged even by
applying the field of 6 T due to the strong inter-band interaction.
The DOS calculated from these subbands are shown in Fig. 
\ref{fig:bulk_band}(c) where we applied a Lorentzian smearing of $\eta = 5$ 
meV.
There are general similarities between the experimental data (Fig. 
\ref{fig:kish}) and the theoretical DOS, for instance the existence of 
field independent peak just above $E_F$ and the asymmetry of peak 
structures above and below $E_F$.
However, we also notice several quantitative disagreements:
First, the peak energy separations in the theoretical DOS are much smaller
than those observed experimentally.
Second, the relative height of the field independent peak to the others is
much smaller in the theory.

These discrepancies can be resolved by calculating the LDOS
at the surface layer of bulk graphite.
We decomposed the SWMcC Hamiltonian into a Hamiltonian for successive
layers and calculated the surface LDOS by the Green's functions method
\cite{SurfGreen1,tagami}.
The result of the calculation with the surface potential ($\phi_s$) of
$+17$ meV is
shown as the solid line in Fig. \ref{fig:kish}(b) for the case of $B = 6$ T.
The agreement between the experiment and theory is now excellent.
This means that what is measured with STS is the surface LDOS reflecting
the bulk electronic state behind.
This was confirmed by the fact that peak patterns calculated for
underlayers rapidly change and approach that for bulk graphite shown in
Fig. \ref{fig:bulk_band}(c).
The physical meaning of $\phi_s$ is band bending near the surface.
Its possible origin is a local electrostatic potential induced by the tip.
In the calculations, we added $\phi_{s}$ to on-site atomic levels at the
surface layer in the Hamiltonian matrix.
Note that the surface potential affects mostly the $n = 0, -1$ LLs and not
the larger $n$ LLs (see the dashed line in Fig. \ref{fig:kish}(b)).

\begin{figure}[h]
\includegraphics[width = 6.6 cm]{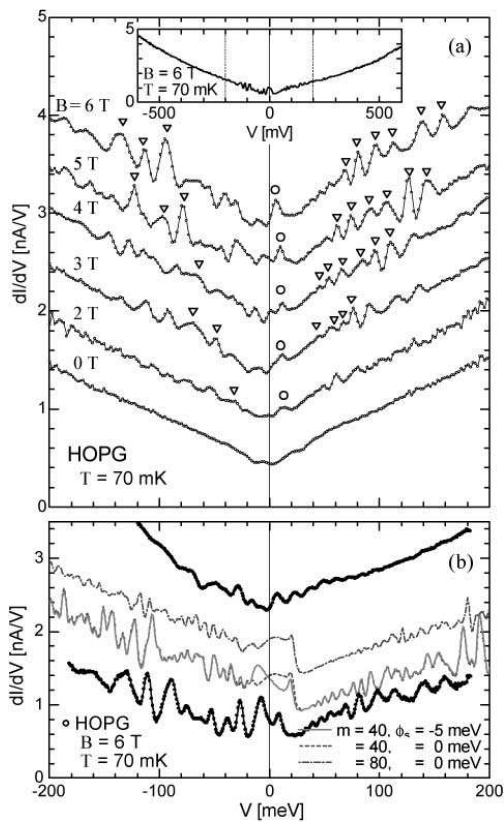}
\caption{\label{fig:HOPG} (a) Tunnel spectra measured at $T=70$ mK for
HOPG in several different fields ($V= 0.2$ V, $I$ = 0.24 nA). Each spectrum
is vertically shifted by 0.5 nA/V for clarity.
The inset shows the spectrum at $B$ = 6 T in a wider energy range.
(b) Comparisons between the calculated LDOS ($\eta = 5$ meV) for the
surface layer of graphite with $m = 40$ layers with (solid line) and
without (dashed line) the surface potential ($\phi_{s} = -5$ meV) and the
measured tunnel spectrum for HOPG at $B=6$ T (dots; $V= 0.2$ V, $I$ = 0.24
nA). The dash-dotted line is the calculation for $m = 80$ layers without
$\phi_{s}$.
The topmost spectrum shows experimental data ($V= 0.2$ V, $I$ = 0.24 nA) 
for an HOPG sample with a smaller $\alpha$ value ($= 1.9 \times 10^{3}$) 
than that for the other data.}
\end{figure}

Next we show tunnel spectra obtained for the HOPG sample in Fig. 
\ref{fig:HOPG}(a).
Each spectrum is an average of the raw data at 100 different positions over
$90 \times 42$ nm$^2$ area.
Again, we saw a series of growing and separating peaks with increasing
field as well as a nearly field independent peak just above $E_F$.
The peak structures are, however, much more pronounced and complicated than
those for Kish graphite except for the $n = 0, -1$ LLs.
The peak amplitudes are comparable to those observed in the 2DES at the
Fe/InAs(110) surface \cite{morgenstern}.
All these indicate that in HOPG an effectively 2D electronic state is
created unlike in Kish graphite.

The somewhat irregular LL peak pattern for HOPG is explained by calculated 
LDOS at the surface of graphite with finite layer thickness ($l^*$).
We calculated the LDOS at each atomic site and each layer by diagonalizing
the decomposed Hamiltonian matrices.
As can be seen in Fig. \ref{fig:HOPG}(b), the experimental data for HOPG
at $B =$ 6 T are
reproduced reasonably well by the calculation for $m =$ 40 layers, where
$m$ is the number of graphene layers ($m \equiv l^{*}/0.335$ nm).
We assumed here $\phi_{s}= -5$ meV.
The many irregular peaks are results of complicated phase interference
between the incoming and outgoing wave functions at the surface.
The calculations show that the peak amplitudes at high energies decrease
with increasing $m$ (cf. the dash-dotted line for $m =$ 80 in Fig. 
\ref{fig:HOPG}(b)) and that, of course, the peak pattern approaches the 
result for bulk
surface as $m \rightarrow \infty$ \cite{tagami}.
It is also shown that the peak structure near $E_F$ due to the $n = 0, -1$
LLs does not depend appreciably on $m$.

The electronic identification of HOPG as graphite with finite thickness is
consistent with the fact that it contains a much higher stacking fault
density than Kish or natural graphite.
This fact was checked by measuring the ratio ($\alpha$) of the in-plane
electrical conductivity to the out-of-plane one.
The $\alpha$ values for Kish graphite and HOPG are 190 and $3.1 \times
10^{3}$, respectively.
Furthermore, another HOPG sample with a lower $\alpha$ value shows less
corrugated tunnel spectra (the topmost spectrum in Fig. \ref{fig:HOPG}(b)) 
in accordance
with the above mentioned calculations.

In the following, we discuss about the origin of $\phi_{s}$.
Experimentally, although the tunnel spectra within $\pm$ 30 meV of $E_F$
changed significantly depending on the tip condition, those at higher
energies were always unaffected.
In addition, the overall spectra were stable for many hours unless we
changed the tip condition intentionally, for example, by applying high
voltage pulses ($\approx 8$ V).
Thus the origin is attributable to the electrostatic potential caused by
the tip.
Note that, in graphite, the DOS in the vicinity of $E_F$ should be very
sensitive to $\phi_s$, since the $n = 0, -1$ subbands are much narrower
compared to the others (see Figs. \ref{fig:bulk_band}(a)(b)).
In other words, the LDOS peak associated with the $n = 0, -1$ LLs can
easily be enhanced even by a small amount of $\phi_s$ like in Fig. 
\ref{fig:kish}.
This is also seen in Figs. \ref{fig:kish}(b) and \ref{fig:HOPG}(b) where 
the calculated LDOS peaks near $E_F$ are quite different between with and 
without the small surface potentials,
whilst little is different for the higher energy peaks.
We did not see a clear correlation between $\phi_{s}$ and the tip material
(W or PtIr) which may suggest the importance of the work function difference
between the tip and sample \cite{wildoer}.
The relatively small effects on the LL structure (except for the $n = 0,
-1$ LLs) caused by the tip compared to the 2DES in semiconductors are
presumably due to the higher carrier densities and, hence, more effective
screening effect in graphite.

\begin{figure}
\includegraphics[width = 8.5 cm]{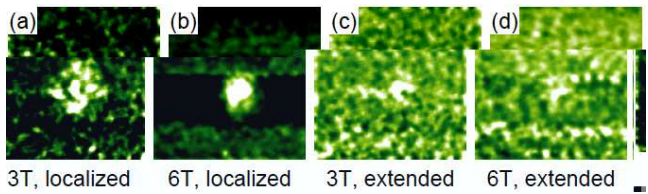}
\caption{\label{fig:HOPG_defect} $dI/dV$ images ($80 \times 80$
nm$^2$, $I =$ 0.2 nA) near point defects at the HOPG surface at $T =$ 30
mK. (a) $V = 30$ meV, $B = 3$ T and (b) 36 meV, 6 T corresponding to
energies in between $n =$1 and 2 LLs. (c) 33 meV, 3 T and (d) 43 meV, 6 T
corresponding to energies of the $n =$2 LL.
The contrast in all images is identical.}
\end{figure}

We observed a circular distribution of the LDOS around point defects on the
HOPG surface in the $dI/dV$ image at an energy in between $n =$1 and 2 LLs
at $B =$ 3 T as shown in Fig. \ref{fig:HOPG_defect}(a).
The radial extension is roughly equal to the magnetic length,
$\sqrt{\hbar/eB}$ ($= 15$ nm at 3 T), of the 2DES, and decreases as $B$
increases to 6 T (see Fig. \ref{fig:HOPG_defect}(b)) suggesting 
localization of the electronic
wave function along an equipotential contour around the defects.
In contrast, the $dI/dV$ images at energies corresponding to the $n =$ 2 LL
do not show such a distribution but complicated nonlocal networks of the
wave function in addition to moderately bright local spots at different
positions from the defects (Figs. \ref{fig:HOPG_defect}(c)(d)).
These observations \cite{defect} indicate the existence of the 2DES
in HOPG and likely the drift- and extended states previously
observed in the Fe/InAs(110) system \cite{morgenstern}.

It is generally expected for the LL structure to have a rather strong
atomic-site dependence.
Actually, the calculated LDOS peak structures are qualitatively different
between the A and B sites in the case of graphite with finite thickness
\cite{tagami}.
Note that the calculated LDOS shown in Fig. \ref{fig:HOPG} is the sum of 
LDOS on the two sites.
However, experimentally, averaged tunnel spectra on the B sites differ from
those on the A sites
only within the experimental errors.
We speculate that the expected site dependence was masked by
dominating tunnel currents
from neighboring B-site atoms even when the tip is just above an A-site atom.

In conclusion, we succeeded in observing clearly the Landau quantization in
bulk material (graphite) with STS in high magnetic fields and very low
temperatures.
By comparing with calculated LDOS, we found that the measured tunnel
spectra for Kish graphite and HOPG well represent the LDOS at the surface
layers of bulk graphite and graphite with finite thickness of about 40
layers, respectively.
The more pronounced LL structure in HOPG indicates the much stronger
2D nature of the electronic state than previously thought.
Our results provide a spectroscopic statement that HOPG has a high-mobility
quasi 2DES which is eligible to show the quantum Hall effect as was
conjectured by the recent transport measurements \cite{kopelevich}.
This was supported by the wave function mappings near surface defects which
show the localization and extension of the electronic state depending on
energy relative to the LL structure.
The present work is the first example to extract quantitative
information on the electronic states in bulk or finite thickness layered
material from the surface LDOS measured with STS.
This further expands the potentialities of the STS technique for material
science.

We thank H. Aoki and Y. Takada for valuable discussions.
We are also grateful to Y. Iye for kindly supplying us the Kish graphite
samples.
This work was financially supported by Grant-in-Aid for Scientific Research
from MEXT, Japan and ERATO project of JST.
T.M. and Y.N. acknowledge the JSPS Research Fellowship for Young
Scientists.

\end{document}